\documentclass[twocolumn,superscriptaddress,showpacs,preprintnumbers,
               amsmath,amssymb]{revtex4}

\usepackage{graphicx}   
\usepackage{dcolumn}    
\usepackage{bm}         
\usepackage{color}      

\def\mem#1#2#3{  \left\langle #1 \left\vert  #2 \right\vert #3 \right\rangle   }

\def\etal{\textit{et al.{}}}

\begin{document}


\title{Bound-Free Electron-Positron Pair Production in Relativistic 
       Heavy-Ion Collisions}

\author{M.~Y.~\c{S}eng\"{u}l}
\email{myilmaz@khas.edu.tr}
\affiliation{\.{I}stanbul Technical University, Faculty of Science and Letters,
             34469, \.{I}stanbul,Turkey}
\affiliation{Kadir Has University, Faculty of Science and Letters,
             34083 Cibali, Fatih-\.{I}stanbul,Turkey}

\author{M.~C.~G\"{u}\c{c}l\"{u}}
\affiliation{\.{I}stanbul Technical University, Faculty of Science and Letters,
             34469, \.{I}stanbul,Turkey}

\author{S. Fritzsche}
\affiliation{Department of Physical Sciences, P.O.~Box 3000,
             Fin-90014 University of Oulu, Finland}%
\affiliation{GSI Helmholtzzentrum f{\"u}r Schwerionenforschung, 
             D-64291 Darmstadt, Germany}%

\date{\today}


\begin{abstract}
The bound-free electron-positron pair production is considered for 
relativistic heavy ion collisions. In particular, cross sections are 
calculated for the pair production with the simultaneous capture of the 
electron into the $1s$ ground state of one of the ions and for energies 
that are relevant for the Relativistic Heavy Ion Collider (RHIC) and the 
Large Hadron Colliders (LHC). In the framework of perturbation theory, 
we applied Monte-Carlo integration techniques to compute the lowest-order
Feynman diagrams amplitudes by using Darwin wave functions for the bound 
states of the electrons and Sommerfeld-Maue wave functions for the 
continuum states of the positrons. Calculations were performed 
especially for the collision of $Au + Au$ at 100 GeV/nucleon and 
$Pb + Pb$ at 3400 GeV/nucleon.
\end{abstract}

\pacs{25.75.Dw; 25.30.Rw}    
\keywords{Pair Production, Electron Capture, QED, Monte Carlo Method}
\maketitle

\section{\label{sec:level1}Introduction}\label{s1}

The bound-free electron-positron pair production plays an important role 
at modern colliders such as the Relativistic Heavy Ion Collider (RHIC) or
the Large Hadron Colliders (LHC), since it may restrict the luminosity
of the ion beams that will be available. Especially in peripheral collisions
of the ions, it is known that a large number of lepton pairs can be produced 
owing to the Lorentz contracted electromagnetic fields that occur in course 
of the collisions. In the bound-free pair production, the electron is 
captured by one of the colliding ions
\begin{eqnarray}
\label{BFPP-process}
   Z_a+Z_b &\rightarrow& (Z_a+e^-)_{1s_{1/2,\ldots}} \:+\: Z_b \:+\: e^+
\end{eqnarray}
and leads to the loss of the (one-electron) ion from the beam. The bound-free 
pair production (BFPP) will therefore be an important problem at the LHC;
in fact, this process does not only reduce the intensity of the beam but also
leads to a separate beam of one-electron ions that strikes the beam-pipe 
about 140 meters away from the interaction point. In the worst scenario, there
might be enough energy in this separated beam to quench the LHC magnets as it
was first pointed out by Bruce \etal{}\cite{let3} but was investigated in further detail
in Refs.~\cite{nuc1, let1, let2}.

A first computation on the bound-free pair production cross sections
were performed by Baltz, Rhoades-Brown and Weneser \cite{a1} in the mid 
1990ies.  These authors used large-basis coupled-channel Dirac-equation of BFPP in
their calculations. 
In particular, Baltz and coworkers derived a simple formula
for the BFPP cross sections at ultra-relativistic energies ($\gamma$ = 23 000)
\begin{eqnarray}\label{e111}
   \sigma_{BFPP}=Aln\left(\gamma\right)+B
\end{eqnarray}
where $A$ and $B$ are the parameters independent of energy. In this expression for
the cross section, the $Aln\gamma$ term represents the region of large 
impact parameters, and $A$ was calculated by using perturbation theory, while
the parameter $B$ represents the contributions from small impact parameter
and includes both, perturbative and non-perturbative parts. The production of
bound-free electron-positron pairs was calculated also by Bertulani and 
Baur \cite{rep2} who used a semi-classical method in order to calculate the
production of bound-free pairs at energies and for collision systems 
appropriate for the RHIC facility. From these computations it was found that
the BFPP cross sections for the capture of the electron into the
$ns$ excited states of the ion decreases with $\approx1/n^3$, which means a
factor of 1/8 for the $L$-shell and to a net effect of all $ns$ excited states
of approximately 20~\%{} in total \cite{rep1, a2}.

An alternative method was later applied by Rhoades-Brown and coworkers 
\cite{a5} who performed an "exact" integration of the Feynman diagrams by using
Monte-Carlo techniques. In this work, the cross section for the capture of an
electron was obtained as the convolution of the amplitude of $direct$ and $crossed$ Feynman
diagrams in Fig.~\ref{f2},
\begin{eqnarray}\label{e122}
   B(k,q;\mathbf{p_{\bot}}) & = &
   A^{(+)}(k,q;\mathbf{p_{\bot}})  
   \nonumber \\
   &  & \quad
   \:+\: A^{(-)}(k,q;\mathbf{k_{\bot}+q_{\bot}-p_{\bot}}) \, ,
\end{eqnarray}
with the momentum distribution of the bound-state wave function. The results 
from these Feynman-Monte Carlo computations were compared with the 
Weizsacker-Williams calculations by Baur and Bertulani \cite{rep2} and were 
found larger by about a factor of 3, a discrepancy which was explained later 
in a comment by Baur~\cite{a8}. In the present work, we have calculated 
the cross section for the capture of an electron into the $K$-shell by applying
a Monte-Carlo integration for the lowest-order Feynman diagrams as shown in
Fig.~\ref{f2}. This procedure is known also in the literature \cite{d2,a6}
as the two-photon method since the colliding nuclei (nucleus $a$ and nucleus $b$)
exchange one photon (total two photon) and the two-photon-exchange diagrams
are proportional to $Z\alpha$.
In contrast to our previous 
computations, where plane-waves were applied for both the electron and the
positron  \cite{d2,a6}, we here apply bound $K$-shell 
wave-functions for the electron as well as modified plane-waves functions for 
the positrons that includes a correction due to the distortion by the 
`bound' electron. In fact, this distortion of the positron wave function 
arises from the necessary (re-) normalization of the continuum waves in order
to account for the reduction of the wave functions of the positron near 
to the nucleus to which the electron is localized \cite{nucpa}. In the
literature, these (one-particle) functions are known as Sommerfeld-Maue wave 
functions for the positrons and Darwin wave functions for the bound electrons. 
Similar wave functions have been applied also in Refs.~\cite{rep2, nucpa} 
for studying the captured electrons and free positrons.

In the next section, we first present the formalism for evaluating the pair 
production cross sections with an electron bound to one of the ions. Apart 
from the representation of the electron and positron states, this includes 
the analysis and a step-wise simplification of the bound-free amplitudes by 
using the wave functions from above. In Section~III, then, the differential 
BFPP cross sections are calculated as function of the transverse and
longitudinal momentum, the energy and rapidity, and especially for those
collision energies of the ions that are relevant for the RHIC and LHC 
facilities. A comparison of our Monte Carlo-Feynman calculations with previous
computations is made. Finally, a few conclusions are drawn in Section~IV.

\section{Theoretical background}

\begin{figure}
\includegraphics{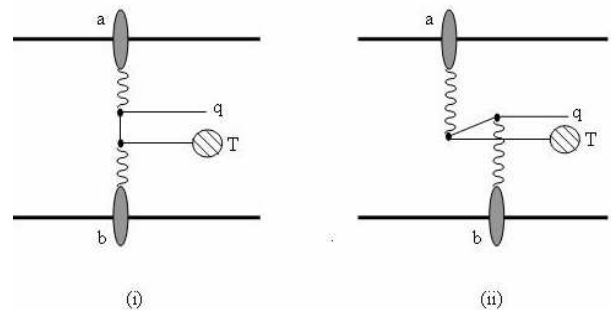} 
\caption{\label{f2} Lowest-order Feynman diagrams for the pair production of a
bound-free electron-positron pair in heavy-ion collisions: 
(i) \textit{direct} and (ii) \textit{crossed} diagrams for the simultaneous 
capture of the electron into a bound state of target (T) ion. 
In the figure, $a$ and $b$ represents the two ions, and $\mathbf{q}$ is the 
momentum of the positron \cite{a5}. }  
\end{figure}

Lowest-order perturbation theory in the framework of quantum electrodynamics
(QED) has been applied to derive and calculate the cross section for
generating bound-free electron-positron pairs in relativistic heavy-ion 
collisions. For this electron-positron BFPP process, the signature is that 
the electron is captured by one of the colliding ions, while the \textit{free} 
positron leaves the collision system. In lowest QED order, this process is 
described by the two Feynman diagrams, the (so-called) \textit{direct} and 
\textit{crossed} terms, as depicted schematically in Fig.~\ref{f2}. These 
diagrams represent the leading contributions to the bound-free pair production 
as appropriate especially  for the high collision energies 
available at the RHIC and LHC facilities.

For two ions $a$ and $b$, that collide with high energy, the leading 
contributions to the pair production arise from those Feynman diagrams for 
which each ion interacts exactly once with the electromagnetic field of the other ion. 
This restriction gives rise to the 
\textit{direct} and \textit{crossed} diagrams from Fig.~\ref{f2}, and to an
gauge-invariant total amplitude that is Lorentz covariant. In the sudden (or 
impulse) approximation, the pair creation with simultaneous capture of the 
electron by one of the ions is then described by convoluting the electron line 
in Fig.~\ref{f2} with the momentum wave function of the final bound state 
\cite{a5}. In the following, we consider the collision of the ions in the
`collider frame' with zero total momentum of the overall system.  
In these coordinates, the two nonzero components of 
the vector potential of ion $a$ is given by
\begin{subequations}\label{e2}
\begin{eqnarray}\label{e2a}
   A_{a}^0 & = & -[2\pi Ze] \, \delta (q^{'}_{0}) \, 
                 e^{[-i\mathbf{q}_{\perp}\cdot\frac{\mathbf{b}}{2}]}
   \int^{\infty}_{-\infty} d^3 \mathbf{r} \:
   \frac{e^{-i\mathbf{q^{'}}\cdot\mathbf{r}}}{\left|\mathbf{r}\right|} 		
\end{eqnarray}
\begin{eqnarray}\label{e2b}
   A_{a}^z & = & \beta A_{a}^0 \, ,
\end{eqnarray}
\end{subequations}
while, similarly, the potential of ion $b$ is obtained by just substituting
the impact parameter $b\rightarrow -b$ in the exponent of Eq.~(\ref{e2a}) 
and the relative velocity $\beta \equiv \frac{v}{c} \rightarrow -\beta$ in 
the formulas above. In these expressions, moreover, $v$ refers the 
velocity of nucleus $a$ moving from right to left parallel to the $z$ axis.
Nucleus $b$ moves from left to right with velocity $-v$. In this work,
we consider only the symmetric collisions (equal mass and equal charge)
of the heavy ions, although the calculations can be done easily for the 
asymmetric collisions. For the overall collision system, therefore, the classical 
four-potential $A^\mu$ can be written as \cite{d2, a6}:
\begin{eqnarray}\label{e3}
   A^\mu & = & A_{a}^\mu \,+\, A_{b}^\mu 
\end{eqnarray}
which describe a retarded Lienard-Wiechert interactions.

To evaluate the diagrams in Fig.~\ref{f2}, we need of course a proper set of
one-particle states in order to represent the electron-positron pairs after 
their generation in the field of the moving ions. For the outgoing positron,
the spinor structure is
\begin{eqnarray}\label{e3a}
   \mathbf{u}^{(+)}_{\sigma_q} & = &
   \sqrt{\frac{E^{(+)}_{q}+mc^{2}}{2mc^{2}}} \,
   \left[\begin{array}{c}
      \phi^{(s)} \\[0.1cm]
      \frac{\bm\sigma\cdot\mathbf{p}c}{E^{(+)}_{q}+mc^2}\phi^{(s)}
   \end{array} \right]
\end{eqnarray}
(for spinors with positive energy $E^{(+)}_{q}>0$), and where 
$\phi^{(s)}=\chi^{(s)}_{1/2}$ denotes a Pauli spinor and $s=\pm1/2$ its spin 
projection. Here, after the creation of electron-positron pairs, electron is captured
by one of the colliding ions and positron becomes free which is described by the plane-waves
\begin{eqnarray}\label{e7}
   \Psi^{(+)}_{q} & = & 
   N_{+}\left[e^{i\mathbf{q}\cdot\mathbf{r}}\,
   \textbf{u}^{(+)}_{\sigma_q} \,+\, \Psi^{'}\right],
\end{eqnarray}
and together with the (correction) term $\Psi^{'}$ in order to account for 
the distortion due to the charge of one of the nucleus. In expression (\ref{e7}), moreover, 
\begin{eqnarray}
   N_{+} & = & e^{-\pi\!a_{+}/2} \, \Gamma(1+ia_{+})\, , \qquad
   a_{+}=\frac{Z\!e^{2}}{v_{+}} \, ,
\end{eqnarray}
is a normalization constant which accounts for the distortion of the wave 
function is acceptable for $Z\alpha \ll 1$ \cite{rep2,nucpa, epl},
and where $\alpha \,=\, e^{2}/\hbar c \,\cong\, 1/137$ is the fine structure 
constant and $v_{+}$ the velocity of the positron in the rest frame of the 
ion, into which the electron is captured in course of 
process~(\ref{BFPP-process}). For sufficiently large energies of the ions, 
we can approximate $v_{+} \,\cong\, c=1$ by using natural units 
($\hbar=c=m=1$), and which are used throughout this work if not stated 
otherwise.

After the pair production~(\ref{BFPP-process}) has occurred, the electron is 
captured by one of the ions and, thus, need to be described as a bound state. 
In a semi-relativistic approximation, these electron states are often 
represented by \cite{book1, book2}
\begin{eqnarray}
   \Psi^{(-)} & = &
   \left( 1-\frac{i}{2m} \, \bm\alpha\cdot \bm\nabla \right) \:
   \textbf{u} \, \Psi_{\rm\, non-rel} (r) \, ,
\end{eqnarray}
i.e.\ in terms of the non-relativistic (ground) state function
\begin{eqnarray} \label{e8}
   \Psi_{\rm\, non-rel}(r) & = & 
   \frac{1}{\sqrt{\pi}} \left( \frac{Z}{a_H} \right)^{3/2} \, e^{-Zr/a_H},
\end{eqnarray}
of the hydrogen-like ion, and where $\textbf{u}$ represents the spinor part 
of the captured electron and $a_H=1/e^2$ the Bohr radius of atomic hydrogen.

Using the positron and electron states from above, the direct diagram in 
Fig.~\ref{f2}(i) can be written as: 
\begin{widetext}
\begin{eqnarray} \label{d11}
   \mem{\Psi^{(-)}}{\,S_{ab}\,}{\Psi^{(+)}_q} & = &
   i\,\sum_{p}\sum_{s} \int^{\infty}_{-\infty} \frac{d\omega}{2\pi} \:
   \frac{\mem{\Psi^{(-)}}{V_a(\omega-E^{(-)})}{\chi^{(s)}_{p}} \,
         \mem{\chi^{(s)}_{p}}{V_b(E^{(+)}_{q}-\omega)}{\Psi^{(+)}_{q}}
       }{(E^{(s)}_{p}-\omega)}
   \\[0.2cm]
\label{e9}
   & = & i\,\sum_p\sum_s 
   \int^{\infty}_{-\infty} \frac{d\omega}{2\pi} 
   \int^{\infty}_{-\infty} d^3\textbf{r} 
   \left(1+\frac{i}{2m} \bm\alpha\cdot\bm\nabla\right) \:
   \Psi_{\rm\, non-rel}(r) \: e^{i\mathbf{p}\cdot\mathbf{r}} \:
   A_a(\mathbf{r}; \omega-E^{(-)})  
   \nonumber \\[0.1cm]
   &   & \hspace*{-0.6cm} \times 		   
   \int^{\infty}_{-\infty} d^3\textbf{r}^{'} \, N_{+} \, 
   e^{-i(\mathbf{p}-\mathbf{q})\cdot\mathbf{r}^{'}}
   A_b(\mathbf{r}^{'}; E^{(+)}_{q}-\omega) \,
   \frac{\mem{\textbf{u}}{(1-\beta\alpha_z)}{\textbf{u}^{(s)}_{{\sigma_p}}}
         \mem{\textbf{u}^{(s)}_{\sigma_p}}{(1+\beta\alpha_z)}{
         \textbf{u}^{(+)}_{{\sigma_q}}}}{(E^{(s)}_{p}-\omega)} \, ,
\end{eqnarray}
\end{widetext}
and where $V_a$ and $V_b$ are the potentials of the two nuclei 
$a$ and $b$, 
\begin{subequations}
\begin{eqnarray}
   V_a & = & (1-\beta\alpha_z) \: A_a^0
   \\[0.1cm]
   V_b & = & (1+\beta\alpha_z) \: A_b^0 \, ,
\end{eqnarray}
\end{subequations}
respectively, and with the vector potentials taken from above 
[cf.~Eq.~\ref{e2}].
Apparently, therefore, the overall (direct) amplitude contains a three-fold 
integration over the coordinates of the bound electron ($d^3 \mathbf{r}$), 
the coordinates of the free positron ($d^3 \mathbf{r'}$) as well as 
the integration over the frequency $\omega$ of the virtually exchanged
photons between the heavy ions. In the evaluation of Eq.~(\ref{d11}),
moreover, we have used the completeness relation
\begin{eqnarray} \label{aa1}
  \sum _{p}\left|\chi^{(+)}_{p}\right\rangle\left\langle 
                 \chi^{(+)}_{p}\right|+\left|\chi^{(-)}_{p}
	   \right\rangle\left\langle \chi^{(-)}_{p}\right| & = & 1
\end{eqnarray}
with $\chi^{(s)}_{p}$ being:
\begin{subequations}
\begin{eqnarray}
   \chi^{(+)}_{p} & = & 
   e^{i \mathbf{p}\cdot\mathbf{r}} \: \textbf{u}^{(+)}_{\sigma_p}
   \\[0.2cm]
   \chi^{(-)}_{p} & = &
   e^{-i \mathbf{p}\cdot\mathbf{r}} \: \textbf{u}^{(-)}_{\sigma_p} \, .
\end{eqnarray}
\end{subequations}
In this notation of the one-particle states, again, 
$\textbf{u}^{(s)}_{\sigma_p}$ refers to the spinor part of the intermediate 
state of $\chi^{(s)}_{p}$, and the summation over the spin 
and momentum of the one particle can be replaced by
\begin{eqnarray} \label{eeee8}
   \sum_{p}  & = & \sum_{\sigma_p}\sum_{\mathbf{p}}\rightarrow 
   \sum_{\sigma_p}\int\frac{d^3\mathbf{p}}{(2\pi)^3} \, .
\end{eqnarray}

We can perform the integration over $\mathbf{r}$ and $\mathbf{r}'$ explicitly
in Eq.~(\ref{e9}). If we first consider the integral over $\mathbf{r}$, the
two parts of this integral

\newpage
\begin{widetext}
\begin{eqnarray}\label{e10}
   &   & \hspace*{-1.1cm}
   \int^{\infty}_{-\infty} d^3\mathbf{r} \, 
   \left( 1+\frac{i}{2m} \bm\alpha\cdot\bm\nabla \right) \:
   \Psi_{\rm\, non-rel}(r) \: e^{i\mathbf{p}\cdot\mathbf{r}} \,
   A_a(\mathbf{r}; \omega-E^{(-)}) 
   \nonumber \\[0.3cm]
   & = &
   \int^{\infty}_{-\infty} d^3\mathbf{r} \, 
   \Psi_{\rm\, non-rel}(r) \: e^{i\mathbf{p}\cdot\mathbf{r}} \,
   A_a(\mathbf{r}; \omega-E^{(-)}) \;+\;
   \left(\frac{i}{2m}\right) \, 
   \int^{\infty}_{-\infty} d^3\mathbf{r} \, \bm\alpha\cdot\bm\nabla\,
   \Psi_{\rm\, non-rel}(r) \: e^{i\mathbf{p}\cdot\mathbf{r}} \,
   A_a(\mathbf{r}; \omega - E^{(-)}) 
\end{eqnarray}
can be analyzed independently. Making use of the explicit form of the
non-relativistic $1s-$function $\Psi_{\rm\, non-rel}(r)$ and the vector 
potential $A_a(\mathbf{r};\omega-E^{(-)})$, we can write the first part
(on the rhs) of Eq.~(\ref{e10}) as 
\begin{eqnarray}\label{e11}
   \int^{\infty}_{-\infty} d^3\mathbf{r} \, 
   \Psi_{\rm\, non-rel}(r) \: e^{i\mathbf{p}\cdot\mathbf{r}} \,
   A_a(\mathbf{r}; \omega-E^{(-)}) & = & 
   - [2\pi Ze] \, \delta (p^{'}_{0}) \: 
   e^{\left[-ip^{'}_{y}\frac{b}{2}\right]} \,
   \frac{1}{\sqrt{\pi}} \left(\frac{Z}{a_H}\right)^{3/2}
   \int^{\infty}_{-\infty} d^3\mathbf{r} \, e^{-Zr/a_H} \,
   \frac{e^{i\mathbf{p}\cdot\mathbf{r}}}{\mathbf{r}} \, 
   \nonumber \\[0.2cm]
\label{e13}
   & = & 
   - \frac{8\pi^2 Ze}{ \sqrt{\pi} } \,
   \left(\frac{Z}{a_H}\right)^{3/2} \,
   \frac{e^{\left[i\mathbf{p}_{\bot}\cdot\frac{\mathbf{b}}{2}\right]}}{
         \left(\frac{Z^2}{a^{2}_{H}} + \frac{p^{2}_{z}}{\gamma^2} +
	       \mathbf{p}\:^{2}_{\bot}\right)} \,
   \delta(\omega-E^{(-)}-\beta p_z) \, .
\end{eqnarray}
\end{widetext}
Here, in the second line, we made use of the known integral
\begin{eqnarray}\label{e12}
   \int^{\infty}_{-\infty} d^3\mathbf{r} \, e^{-Zr/a_H} \,
   \frac{e^{i\mathbf{p}\cdot\mathbf{r}}}{\mathbf{r}} 
   & = & \frac{4\pi}{\left(\frac{Z^2}{a^{2}_{H}} \,+\, \mathbf{p}\:^2\right)} 
\end{eqnarray}
together with the Lorentz transformation as displayed in Appendix A.
In Eq.~(\ref{e13}), moreover, $p_0$ represents the energy term, $p_z$ 
the longitudinal momentum and, $p_\bot(p_y)$ is the transverse momentum 
of the intermediate state as given by Eq.~(\ref{aa1}).

The second part of the integral (\ref{e10}) 
\begin{widetext}
\begin{eqnarray}\label{e14}
   &   & \hspace*{-2.5cm}
   \left(\frac{i}{2m}\right)\,
   \int^{\infty}_{-\infty} d^3\mathbf{r} \, \bm\alpha\cdot\bm\nabla\,
   \Psi_{\rm\, non-rel}(r) \: e^{i\mathbf{p}\cdot\mathbf{r}} \,
   A_a(\mathbf{r}; \omega-E^{(-)}) 
   \nonumber \\[0.1cm]
   & = &
   \left(\frac{i}{2m}\right) \frac{1}{\sqrt{\pi}}
   \left(\frac{Z}{a_H}\right)^{3/2}
   \left\{ -[2\pi Ze] \, \delta(p^{'}_{0}) \: 
           e^{\left[-ip^{'}_{y}\frac{b}{2}\right]} 
	   \int^{\infty}_{-\infty} d^3\mathbf{r} \; \bm\alpha\cdot\bm\nabla \;
	   e^{-Zr/a_H} \; \frac{e^{i\mathbf{p}\cdot\mathbf{r}}}{\mathbf{r}}
   \right\} \, 
   \nonumber \\[0.1cm]
   & = &
   -\frac{1}{2m}\,\frac{1}{\sqrt{\pi}}\,
   \left(\frac{Z}{a_H}\right)^{3/2} \, 8\pi^2Ze\:\bm\alpha\cdot\mathbf{p}\: 
   \delta(\omega-E^{(-)}-\beta p_z) \,
   \frac{e^{\left[i\mathbf{p}_{\bot}\cdot\frac{\mathbf{b}}{2}\right]}
       }{\left(\frac{Z^2}{a^{2}_{H}}+\frac{p^{2}_{z}}{\gamma^2}+
               \mathbf{p}\:^{2}_{\bot}\right)} 
\end{eqnarray}
can be evaluated by following similar lines, but it now contains the factor 
$\frac{i}{2m} \bm\alpha\cdot\bm\nabla $ which arises from the wave function 
of the captured electron. This additional `derivative'  with regard to 
the coordinates of the electrons can be removed by an integration by parts,
\begin{eqnarray}\label{e15}
   \int^{\infty}_{-\infty} d^3\mathbf{r} \, \bm\alpha\cdot\bm\nabla \:
   e^{-Zr/a_H} \, \frac{e^{i\mathbf{p}\cdot\mathbf{r}}}{\mathbf{r}}
   & = & -i \bm\alpha\cdot\mathbf{p} \:
   \frac{4\pi}{\left(\frac{Z^2}{a^{2}_{H}}+\mathbf{p}\:^2\right)} \, .
\end{eqnarray}
For the overall integral (\ref{e10}), this gives rise to the expression:
\begin{eqnarray}\label{e17}
   &   & \hspace*{-3.5cm}
   \int^{\infty}_{-\infty} d^3\mathbf{r} \,
   \left(1+\frac{i}{2m}\bm\alpha\cdot\bm\nabla\right) \:
   \Psi_{\rm\, non-rel}(r) \: e^{i\mathbf{p}\cdot\mathbf{r}} \,
   A_a(\mathbf{r}; \omega-E^{(-)})
   \nonumber \\[0.2cm]
   & = &
   -\left[1+\frac{\bm\alpha\cdot\mathbf{p}}{2m}\right]
   8\pi^2 Ze \, \frac{1}{\sqrt{\pi}}
   \left(\frac{Z}{a_H}\right)^{3/2} \,
   \frac{\delta(\omega-E^{(-)}-\beta p_z)}{
         \left(\frac{Z^2}{a^{2}_{H}} + \frac{p^{2}_{z}}{\gamma^2} +
	       \mathbf{p}\:^{2}_{\bot}\right)} \:
   e^{\left[i\mathbf{p}_{\bot}\cdot\frac{\mathbf{b}}{2}\right]} \, ,
\end{eqnarray}
where $E^{(-)}$ is the energy of the captured electron.

Using analogue steps, the integral over $\mathbf{r}'$ in Eq.~(\ref{e9}) 
can be written as:
\begin{eqnarray}\label{e18}
   \int^{\infty}_{-\infty} d^3\mathbf{r^{'}} \,
   N_{+} \, e^{-i(\mathbf{p}-\mathbf{q})\cdot\mathbf{r}^{'}} \,
   A_b(\mathbf{r}^{'}; E^{(+)}_{q}-\omega)
   & = &
   - N_{+} \, 8\pi^2 Ze \gamma^2 \:
   \frac{\delta(E^{(+)}_{q}-\omega-\beta(p_z-q_z))}{
         (p_z-q_z)^2+\gamma^2(\mathbf{p}_\bot-\mathbf{q}_\bot)^2} \;
   e^{i\,(\mathbf{p}_\bot-\mathbf{q}_\bot)\,\cdot\,\frac{\mathbf{b}}{2}} \, ,
\end{eqnarray}
where $E^{(+)}_{q}$ is now the energy of the positron. Thus, by combining 
both integrals in Eq.~(\ref{e10}), we obtain for the direct BFPP amplitude 
the explicit expression:
\begin{eqnarray}\label{e19}
   \mem{\Psi^{(-)}}{S_{ab}}{\Psi^{(+)}_q}
   & = & 
   i N_{+} \sum_s\sum_{\sigma_p}
   \int \frac{d^3\mathbf{p}}{(2\pi)^3} \, 
   \int \frac{d\omega}{2\pi}  \,
   e^{i(\mathbf{p}_\bot-\frac{\mathbf{q}_\bot}{2})\cdot\mathbf{b}} \:
   8\pi^2 Ze \frac{1}{\sqrt{\pi}} \left(\frac{Z}{a_H}\right)^{3/2}
   \frac{\delta(\omega-E^{(-)} - \beta p_z)}{
         \left(\frac{Z^2}{a^{2}_{H}} + \frac{p^{2}_{z}}{\gamma^2} +
	       \mathbf{p}\:^{2}_{\bot}\right)}
   \left[1+\frac{\bm\alpha\cdot\mathbf{p}}{2m}\right] \,
   \nonumber \\[0.1cm]
   &   & \hspace*{0.2cm} \times \,	       
   8\pi^2 Ze \gamma^2 
   \frac{\delta(E^{(+)}_{q}-\omega-\beta(p_z-q_z))}{
         (p_z-q_z)^2+\gamma^2(\mathbf{p}_\bot-\mathbf{q}_\bot)^2} \,
   \frac{\mem{\textbf{u}}{(1-\beta\alpha_z)}{\textbf{u}^{(s)}_{\sigma_p}}
         \mem{\textbf{u}^{(s)}_{\sigma_p}}{(1+\beta\alpha_z)}{
              \textbf{u}^{(+)}_{\sigma_q}}
        }{E^{(s)}_{p}-\omega} \, ,
\end{eqnarray}
\end{widetext}
and where $E^{(s)}_{p}$ is the energy of intermediate state.

The vector $\mathbf{p}$ describes the momentum of 
the intermediate (electron and positron) states in the field of the ion  and can be
decomposed into its transverse and parallel part, 
$\mathbf{p}=\mathbf{p}_\bot+p_z$, relative to the motion of the ions. 
In Eq. \ref{e19}, the two Dirac delta functions gives us the component 
along the velocity of the heavy ions
\begin{subequations}\label{e20}
\begin{eqnarray}\label{e20a}
   p_z=\frac{E^{(+)}_{q}-E^{(-)}+\beta q_z}{2\beta} \, ,			
\end{eqnarray}
in terms of the energies and the longitudinal momentum, $q_z$, of the 
positron. Similarly, we can write the frequency of virtually exchanged photons
as
\begin{eqnarray}\label{e20b}
   \omega=E^{(-)}+\beta p_z=\frac{E^{(-)}+E^{(+)}_{q}+\beta q_z}{2} \, ,
\end{eqnarray}
\end{subequations}

\noindent
owing to the conservation of the momentum that is obtained from the Dirac 
delta functions in Eq.~(\ref{e19}). In contrast to the parallel part, however,
the transverse momentum of the electrons $\mathbf{p}_\bot$ is not fixed by 
the kinematics of the collision partners, but also depends on the momentum 
that is carried by the fields. After integrating the Eq.~(\ref{e19}) over 
$\omega$ and $p_z$, and by inserting the values from Eqs.~(\ref{e20}) into 
Eq.~(\ref{e19}), the transition matrix element for a fixed spin and momentum 
state of the positron as well as for a given intermediate state can be 
expressed as
\begin{widetext}
\begin{eqnarray}\label{e21}
   \mem{\Psi^{(-)}}{S_{ab}}{\Psi^{(+)}_q} & = & 
   \frac{iN_{+}}{2\beta} \, \frac{1}{\sqrt{\pi}} \, 
   \left(\frac{Z}{a_H}\right)^{3/2} \,
   \int \frac{d^2p_\bot}{(2\pi)^2} \, 
   e^{i(\mathbf{p}_\bot-\frac{\mathbf{q}_\bot}{2})\cdot\mathbf{b}} \,
   F(-\mathbf{p}_\bot: \omega_a) \,
   F(\mathbf{p}_\bot-\mathbf{q}_\bot: \omega_b) \,
   \mathcal{T}_q (\mathbf{p}_\bot: +\beta)   \, ,
\end{eqnarray}
\end{widetext}
where $\mathbf{b}$ is again the impact parameter of the ion-ion collision, 
and the function $F(q,\omega)$ can be described as the scalar part of the field 
associated with the ions $a$ and $b$ in momentum space.
Explicit form of these scalar fields can be written in terms of the corresponding 
frequencies as 
\begin{subequations}\label{e22}
\begin{eqnarray}\label{e22a}
   F(-\mathbf{p}_\bot:\omega_a)=
   \frac{4\pi Ze}{\left(\frac{Z^2}{a^{2}_{H}}\,+\,
                  \frac{\omega^{2}_{a}}{\gamma^2\beta^2}\,+\,
		  \mathbf{p}^{2}_{\bot}\right)} 	
\end{eqnarray}
for the frequency $\omega_a$, and as
\begin{eqnarray}\label{e22b}
   F(\mathbf{p}_\bot-\mathbf{q}_\bot:\omega_b) & = &
   \frac{4\pi Ze\gamma^2\beta^2}{
         \left(\omega^{2}_{b} + \gamma^2\beta^2(\mathbf{p}_\bot -
	       \mathbf{q}_\bot)^2\right)} \,
\end{eqnarray}
\end{subequations}
for the frequency $\omega_b$, respectively.

Owing to the asymmetry in the behavior of the electrons and positrons        
in course of the BFPP process, the wave functions of the free positron and 
captured electron will differ substantially. This difference gives rise
also to different expressions for the frequencies of the virtual photons 
as emitted by the two nuclei, i.e.\
\begin{subequations}\label{e23}
\begin{eqnarray}\label{e23a}
   \omega_a & = & \frac{-E^{(-)}+E^{(+)}_{q}+\beta q_z}{2}=\beta p_z \, ,
\end{eqnarray}
for ion $a$, and
\begin{eqnarray}\label{e23b}
   \omega_b & = & \frac{E^{(+)}_{q}-E^{(-)}-\beta q_z}{2}=\beta(p_z-q_z) \, .
\end{eqnarray}
\end{subequations}
for ion $b$, respectively. As mentioned before, these frequencies are obtained
from integrating Eq.~(\ref{e19}). Apart from the scalar field of each ion, 
Eq.~(\ref{e21}) contains also the transition amplitudes $\mathcal{T}$ which 
relates the intermediate photon lines to the outgoing electron-positron 
lines. This amplitude depends explicitly on the (relative) velocity of the 
ions $\beta$, the transverse momentum $\mathbf{p}_\bot$, and the momentum of
the positron $q$, and it is given by 
\begin{eqnarray}\label{e24}
   &   & \hspace*{-0.75cm}
   \mathcal{T}_q(\mathbf{p}_\bot:+\beta) 
   \nonumber \\[0.2cm]
   & = & 
   \sum_s \sum_{\sigma_p}
   \frac{1}{\left(E^{(s)}_{p} - \left(\frac{E^{(-)}+E^{(+)}_{q}}{2}\right) 
            -\beta\frac{q_z}{2}\right)} 
   \left[1 + \frac{\bm\alpha\cdot\mathbf{p}}{2m}\right]
   \nonumber \\[0.2cm]
   &   & \times
   \mem{\textbf{u}}{(1-\beta\alpha_z)}{\textbf{u}^{(s)}_{\sigma_p}}
   \mem{\textbf{u}^{(s)}_{\sigma_p}}{(1+\beta\alpha_z)}{
        \textbf{u}^{(+)}_{\sigma_q}}  .
\end{eqnarray}
In this amplitude, moreover, the parallel component of the intermediate state
momentum $p_z$ is determined by Eq. (\ref{e20a}). Finally, let us note 
that the integration over the impact parameter $b$ in Eq.~(\ref{e21}) can be 
carried out also analytically. Following very similar lines, it is possible 
also to evaluate the \textit{crossed}-term amplitude 
$ \mem{\Psi^{(-)}}{S_{ba}}{\Psi^{(+)}_q} $ from Fig.~\ref{f2}.

Having the amplitudes for the \textit{direct} and \textit{crossed} diagram, 
we are now prepared to write down the cross section for the generation of a 
free-bound electron-positron pair in collisions of two heavy ions
\begin{eqnarray} \label{ee8}
   \sigma & = & \int d^2b \: \sum_{q<0} \:
   \left| \mem{\Psi^{(-)}}{S}{\Psi^{(+)}_q} \right|^2 \, ,
\end{eqnarray}
where $S \,=\, S_{ab}+S_{ba}$ denotes the sum of the \textit{direct} and 
\textit{crossed} terms in Fig.~\ref{f2}. Making use of all the simplifications
from above, these cross sections for the  BFPP can be expressed as:
\begin{widetext}
\begin{eqnarray}\label{e25}
   \sigma & = & 
   \int d^2b\sum_{q<0}\left|\left\langle 
   \Psi^{(-)}\left|S_{ab}\right|\Psi^{(+)}_q\right\rangle+\left\langle     
   \Psi^{(-)}\left|S_{ba}\right|\Psi^{(+)}_q\right\rangle\right|^2
   \nonumber \\[0.2cm]	   
   & = &			   
   \frac{\left|N_{+}\right|^{2}}{4\beta^2} \,
   \frac{1}{\pi} \left(\frac{Z}{a_H}\right)^3
   \sum_{\sigma_q} \int \frac{d^3qd^2p_\bot}{(2\pi)^5} \,
   \left(\mathcal{A}^{(+)}(q: \mathbf{p}_\bot) + 
         \mathcal{A}^{(-)}(q: \mathbf{q}_\bot  - \mathbf{p}_\bot)\right)^2 \,,
\end{eqnarray}
with
\begin{subequations}\label{e26}
\begin{eqnarray}\label{e26a}
   \mathcal{A}^{(+)}(q:\mathbf{p}_\bot) & = &
   F(-\mathbf{p}_\bot: \omega_a) \, 
   F(\mathbf{p}_\bot-\mathbf{q}_\bot: \omega_b) \,
   \mathcal{T}_q(\mathbf{p}_\bot:+\beta),				
\end{eqnarray}
and
\begin{eqnarray}\label{e26b}
   \mathcal{A}^{(-)}(q:\mathbf{q}_\bot-\mathbf{p}_\bot) & = &
   F(\mathbf{p}_\bot-\mathbf{q}_\bot: \omega_b) \,
   F(-\mathbf{p}_\bot: \omega_a) \,
   \mathcal{T}_q(\mathbf{q}_\bot-\mathbf{p}_\bot: -\beta) \, .
\end{eqnarray}
\end{subequations}
\end{widetext}
being some proper products of the transition amplitudes and scalar parts 
of the fields as associated with ions $a$ and $b$. All these functions have 
been displayed explicitly in Eqs.~(\ref{e22}) and (\ref{e24}) above. Moreover, 
the square of the normalization constant for the positron wave function is 
equal to \cite{rep2,nucpa}
\begin{eqnarray}\label{e99}
   \left|N_{+}\right|^{2}=\frac{2\pi\!a_{+}}{e^{2\pi\!a_{+}}-1} \, .
\end{eqnarray}
Indeed, this constant appears to be very similar to 
\begin{eqnarray}\label{e98}
   \frac{2\pi\alpha\!Z}{e^{2\pi\alpha\!Z}-1} \, ,
\end{eqnarray}
i.e.~the factor that represents the distortion of the positron wave function
due to the shielding of the nucleus by the electron \cite{nucpa, let4, epl}.

\begin{table}[b]
\caption{Bound-free pair production cross sections $\sigma_{\rm\, BFPP}$
(in barn) for selected collision
systems and cross sections as accessible at RHIC and LHC collider
facilities.} \label{t1}\
\begin{tabular}{l l c c c } \\[-0.1cm]
\hline\hline                \\[-0.3cm]
              &&& This work   & Ref.~\cite{a3} \\[0.1cm]
\hline
RHIC$\quad$     & $Au+Au$ at 100 GeV $\quad$   && 94.5 & 94.9 \\
LHC             & $Pb+Pb$ at 2957 GeV          && 202 & 225   \\[0.1cm]
\hline\hline
\end{tabular}
\end{table}

\begin{figure}[t]
\includegraphics[width=0.47\textwidth]{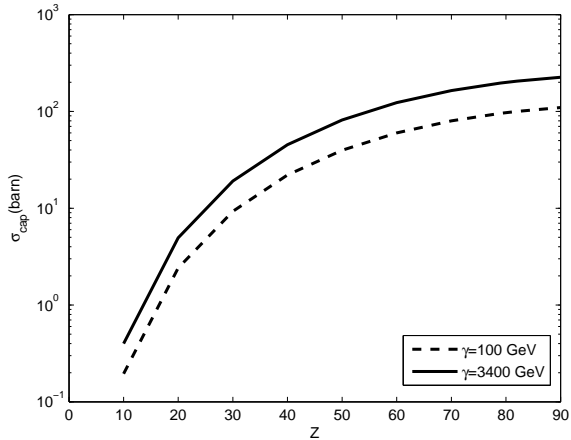}
\caption{BFPP cross sections for two different systems as functions of the nuclear charge $Z$.
BFPP cross sections (in barn) for the symmetric collision of bare ions with 
nuclear charge $Z$ at 100 GeV/nucleon (dashed line) and 3400 GeV/nucleon (solid line). 
The capture cross sections increases by about 3 order of magnitude in going
from $Z = 10$ to $Z = 90$.}
\label{f18}
\end{figure}

\begin{figure}[t]
\includegraphics[width=0.47\textwidth]{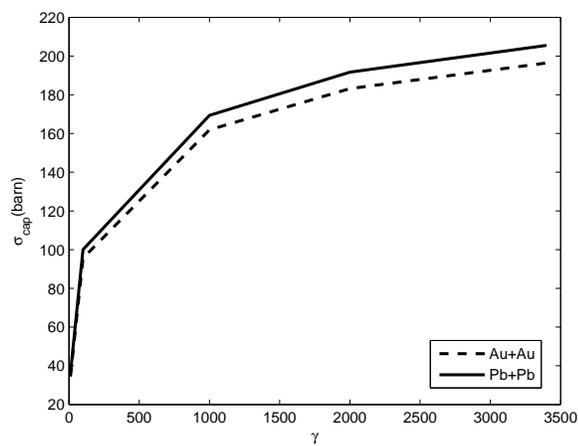}
\caption{BFPP cross sections for two different systems ($Au+Au$-dashed line and $Pb+Pb$-solid line) as functions of 
the $\gamma$. The magnitude of $\gamma$ is going from 10 to 3400.}
\label{f19}
\end{figure}

\section{Results and discussions}\label{s3}

Calculations have been performed for the total production cross sections 
of bound-free electron-positron pairs in relativistic collisions of bare ions.
Theoretical cross sections are obtained especially for the collisions of
$Au + Au$ ions at energies relevant for the RHIC facility as well as for
$Pb + Pb$ ions at LHC energies. These cross sections are compared with those for
the production of free electron-positron pairs. While, however,
the free-pair production includes an eight dimensional integral, the BFPP 
cross sections eventually depend only on five-dimensional integrals. 
To evaluate these amplitudes, Monte-Carlo techniques were utilized, 
and the integrands have been tested on about 10 million randomly chosen
`positions' in order to ensure a sufficient convergence of our theoretical
results. The total numerical errors in the computations is estimated to 
be less or approximately five percent.

As a test of our implementation, first computations of the total BFPP cross 
sections were performed for a few selected collision energies. Table~\ref{t1} 
displays these BFPP cross sections for the two collision systems from above 
and for those collision energies that are relevant for forthcoming experiments
at the RHIC and LHC collider facilities. Our results for the total cross
sections are in good-to-excellent agreement with the previous computations by
Meier \etal{} \cite{a3}, especially for the RHIC energies of 100~GeV per nucleon.
For the much higher collision energies of $\sim 3000$ GeV per nucleon, that will be
available at the LHC storage ring, our theoretical predictions for the BFPP 
cross sections are in contrast lower by about 20~\%{}, compared with the 
computations by Meier and coworkers.

As mentioned above, the correction term $\Psi^{'}$ was omitted to the 
positron wave functions in Eq.~(\ref{e7}), in line with previous experience
and computations of the free (electron-positron) pair production 
for which a perfect agreement with experiment was found by
omitting this term \cite{d2, a6}.  We therefore
conclude that the distortion of the positron states due to the shielding of 
the electron is small and remains negligible for the present computations.

To understand the importance of the bound-free process, Fig.~\ref{f18} displays
the BFPP cross sections for symmetric collisions of ions with charge $Z$ 
as function of the nuclear charge. Cross sections are shown for the two
collision energies $E = 100$ GeV/nucleon (dashed line) and $ 3400$ GeV/nucleon
(solid line) as important for modern accelerators. While the cross sections for
the production of free pair scale approximately
with $\thickapprox (Z\alpha)^{4} ln^3(\gamma)$, the BFPP cross sections increase 
with $\thickapprox (Z\alpha)^{8} ln(\gamma)$. 
In this scaling behavior, the reason for an 
extra factor $Z^{3}$ arises from the bound wave function of the electron, 
while another power in $Z$ comes from the normalization constant for the 
positron wave function.

Fig.~\ref{f19} shows the total BFPP cross sections for two different 
systems as functions of the Lorentz contraction factor $\gamma$.  Our calculation
is done in the center-of-momentum frame, therefore the relationship between the
Lorentz factor $\gamma$ and the collider energy per nucleon in GeV $E/A$ is
given by $\gamma=1/\sqrt{1-v^2}=E/m_{0}$, where $m_{0}$ is the mass of the nucleons.
Results are displayed for $Pb + Pb$ collisions
(solid line) and for $Au + Au$ collisions (dashed line). As function of the 
Lorentz factor $\gamma$, the free pair production scales 
with $ln^{3}(\gamma)$ and the bound-free pair production with $ln(\gamma)$.
All these results are obtained in previous computations
\cite{nucpa, rep2, d1, epjc} and, thus, we also verify that our calculations gives
the similar results.

\begin{figure}[t]
\includegraphics[width=0.47\textwidth]{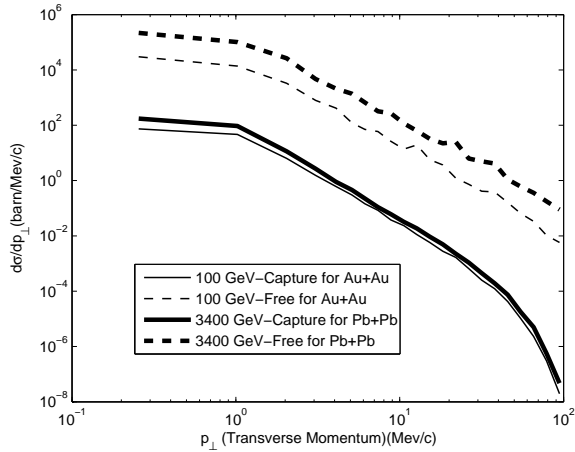}
\caption{The differential cross section as function of the transverse momentum ($p_{\bot}$)
of the produced positrons is shown in the graph.
Calculated differential cross sections are shown for the two collision 
systems $Au+Au$ at 100 GeV per nucleon and $Pb+Pb$ at 3400 GeV per nucleon, respectively. 
When compared with the production of free electron-positron pairs, 
the BFPP cross section for the capture of the electron into the $1s$ ground state
is suppressed by about three orders of magnitude for all transverse momenta 
between 0.1 and 100 MeV/c.}
\label{f5}
\end{figure}

\begin{figure}[t]
\includegraphics[width=0.47\textwidth]{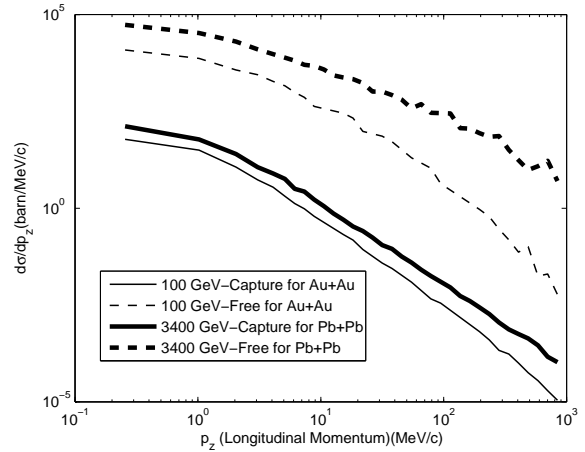}
\caption{The differential cross section is shown as function of the
longitudinal momentum $(p_{z})$ of the produced positrons.
The notations are the same as in Fig.~\ref{f5}.}
\label{f6}
\end{figure}

\begin{figure}[b]
\includegraphics[width=0.47\textwidth]{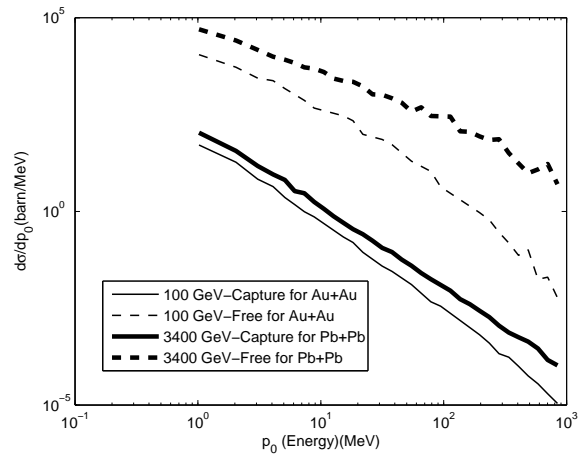}
\caption{The differential cross section as function of the
energy $(p_{0})$ of the produced positrons is shown in the graph. 
The notations are the same as in Fig.~\ref{f5}.}
\label{f7}
\end{figure}

\begin{figure}[t]
\includegraphics[width=0.47\textwidth]{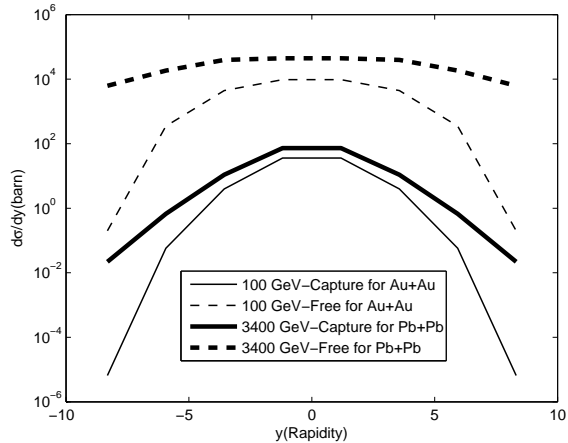}
\caption{The differential cross section is shown as function of the rapidity ($y$).
The notations are the same as in Fig.~\ref{f5}. 
}
\label{f8}
\end{figure}

Results 
for the free and bound-free pair production are displayed in the figures
\ref{f5} to \ref{f8} within the same graph. At collision energies of 100 GeV per nucleon,
as relevant to the RHIC facility, we have considered $Au + Au$ collisions, 
while $Pb + Pb$ collisions were analyzed at 3400 GeV per nucleon, as 
they are hoped to be reached within the near future with the LHC at 
CERN. The same notation is used throughout the figures \ref{f5} to \ref{f8}. 
Thin-lines and thin-dashed 
lines represent the bound-free pair production and free-pair production at 
RHIC collisions of  $Au + Au$, respectively. On the other hand, thick lines and
thick-dashed lines shows the BFPP and free pair production at LHC collisions of
$Pb + Pb$ respectively.

Fig.~\ref{f5} plots the the differential cross section as function of the transverse 
momentum of the produced positrons. From this figure, it becomes clear 
that the bound-free and free pair production distributions display a
rather similar behavior as function of transverse momentum, although the  free-pair distribution function
is larger by about 3 orders of magnitude than the BFPP function.
Obviously, moreover, BFPP distribution function decreases much faster
with the size of the transverse momentum than those for the free-pair 
production.

Figs.~\ref{f6}  and \ref{f7} displays the differential cross sections 
as function of the longitudinal momentum and the energy, respectively. Again,
cross sections are shown for the free and bound-free case. From
Fig.~\ref{f6} we find that the ratio of the RHIC and LHC increases if
the value of the longitudinal momentum increases. Moreover, since the
longitudinal momentum of the positron is much higher than the transverse 
momentum, the energy of the produced positron arises mainly from the 
longitudinal momentum of the positrons. Therefore, the behavior of the differential
cross section as functions of longitudinal momentum and energy
must look very similar as seen from Figs.~\ref{f6} and \ref{f7}. We can 
conclude that energy of the positrons, $E=\sqrt{p^{2}_{\perp}+p^{2}_{z}+1} $, consist
of mainly by the longitudinal momentum of the positrons.

Finally, Fig.\ref{f8} plots  the differential cross section as function of the rapidity.
At RHIC energies, the behavior of the free and bound-free differential cross
sections of rapidity is almost the same, while a different behavior is observed at
LHC energies. For large values of the rapidity, the bound-free cross 
sections decay more rapidly than for the free production sections.
Since the rapidity is a function of the energy and momenta,
\begin{eqnarray}\label{e40}
   y &=& \dfrac{1}{2}ln\left[ \dfrac{p_{0}+p_{z}}{p_{0}-p_{z}} \right] \, ,
\end{eqnarray}
the discrepancies in the behavior in the cross sections as function of the
longitudinal momenta and the energies appears closely related to its the
behavior as function of the rapidity.

\section{Concluding Remarks}\label{s4}

In this work, we have investigated the electron-positron pair production
that is associated with the (simultaneous) capture of the electron into the 
$K-$shell of one of the ions. Calculations of the cross sections have been
performed especially for two collision systems, $Au + Au$ and $Pb + Pb$,
and for energies that are relevant for the RHIC and LHC facilities. In the
framework of QED perturbation theory, the lowest-order Feynman diagrams have
been evaluated by applying Darwin wave functions for the $1s$ bound state
of the electron and Sommerfeld-Maue wave functions for the continuum 
states of the outgoing positron. In line with previous experience, however, 
we have not taken into account the correction term for the positron
wave function whose influence was estimated to be small. We plan to 
incorporate this term in the future to analyze its contribution in further 
detail.

Comparison of our theoretical cross sections is made with previous computations
as far as available. Good agreement is found especially for the total cross
sections of $Au + Au$ collisions at 100 GeV per nucleon as utilized at RHIC. 
For these collision, it is found 
in particular that the free and bound-free pair production cross section
behave very similar as a function of energy with an almost constant factor
of $\backsim 10^3$ with which the bound-free pair production is suppressed.
This is quite different for collision at $\gtrsim 3000$ GeV/nucleon, the
expected conditions at the LHC, where  there is 
discrepancy between BFPP and free pair production differential cross sections
especially for large values of longitudinal momentum, energy and rapidity.

For the total electron-positron pair production, the effect of the `Coulomb 
correction' is known to play an important role \cite{baltz} due to
multi-photon exchange of the produced electron-positron with the colliding nuclei.
This Coulomb correction for the free pair production has a negative value
and it is proportional to $Z^2$ which is obtained by Bethe-Maximon \cite{bethe}. 
So far, these higher-order corrections were not included in the computation
but we plan to derive and calculate this effect in a forthcoming work.

\begin{acknowledgments}
This research is partially supported by the Istanbul Technical University 
and Kadir Has University. We personally thank S.~R.~Klein and G.~Baur 
for valuable advise in calculating the cross sections and M.~\c{S}eng\"{u}l 
for the carefully reading of our article. 
\end{acknowledgments}

\vspace*{0.5cm}
\appendix
\section{Lorentz transformation of the 4-vector potential}

In Section~II, the integrals (\ref{e11} and \ref{e14}) were evaluated by using the
Lorentz-transformed potentials of the Coulomb field of the heavy ions. Here, we derive this
potential in momentum space and perform the Lorentz transformation onto it.
In the rest frame of an ion with the nuclear charge $Z$, fixed to the
coordinates (0,b/2,0), the four-vector potential is given by:
\begin{eqnarray}
   \mathbf{A{'}} & = & 0  \nonumber\\
   A_{0}^{'}     & = & 
   \frac{-Z e}{[x^{'^{2}}+(y^{'}-b/2)^{2}+z^{'^{2}}]^{1/2}} \, .
\end{eqnarray}
In momentum space, this vector potential can be expressed by
\begin{eqnarray}\label{ea2}
   A_{0}^{'}(q^{'}) & = & 
   \int d^{4}x^{'}e^{iq^{'}\cdot x^{'}}A_{0}^{'}(x^{'})
   \nonumber\\
   & = & -[2\pi Ze] \, \delta (q^{'}_{0}) \, e^{[-iq^{'}_{y}\,\frac{b}{2}]} \,
   \nonumber \\[0.1cm]
   &   & \hspace*{1.5cm} \times
   \int^{\infty}_{-\infty} d^3 \mathbf{r^{'}} \:
   \frac{e^{-i\mathbf{q}^{'}\cdot\mathbf{r^{'}}}
       }{\left|\mathbf{r^{'}}\right|} \, ,
\end{eqnarray}
and with 
\begin{eqnarray}
   \mathbf{ r^{'}} & = &
   \mathbf{x^{'}} +(\mathbf{y^{'}}-\mathbf{b^{'}}/2) +\mathbf{z^{'}}.
\end{eqnarray}
Moreover, since $\mathbf{A^{'}} \,=\, 0$ and $A^{'}_{\mu}$ must transforms 
like a four-vector, we obtain
\begin{eqnarray} \label{ea3}
   A_{0}(q) & = &
   \gamma(A^{'}_{0} -\beta A^{'}_{1}) \;=\; \gamma A^{'}_{0}
   \nonumber \\[0.1cm]
   A_{1}(q) & = & \gamma(A^{'}_{1} -\beta A^{'}_{0}) \;=\; 
   -\gamma\, \beta\, A^{'}_{0}
   \nonumber \\[0.1cm]
   \mathbf{A_{\bot}} & = & \mathbf{A^{'}_{\bot}} \;=\; 0 \, .
\end{eqnarray}
Thus, by doing the integration over $\mathbf{r'}$ on the rhs
of Eq.~(\ref{ea2}) and by performing the Lorentz transformations, 
we obtain the potential for the colliding ions:
\begin{eqnarray}\label{ea6}
   A_0 & = & -\left[8\pi^2Ze\right] 
   \delta(q_0+\beta q_1) \, \gamma^2 \,
   \frac{e^{-i\mathbf{q}_{\perp}\cdot\mathbf{b}/2}
       }{\left[q^2_1+\gamma ^2\mathbf{q^{2}_{\bot}}\right]} \, ,
\end{eqnarray}
and where $q_1$ refers to the longitudinal momentum, $\mathbf{q_\bot}$ the transverse momentum and 
$q_0$ is the energy term.

%
%
%
%
%
%
%

\end{document}